\documentclass[a4paper,10pt]{article}
\usepackage{graphicx}
\usepackage{amssymb}
\usepackage{latexsym,cite}
\usepackage{epsfig}
\usepackage{psfrag}
\usepackage{amsmath, bbm, color}

\newcommand{\eq}{\begin{equation}}
\newcommand{\en}{\end{equation}}
\newcommand{\ear}{\begin{eqnarray}}
\newcommand{\rae}{\end{eqnarray}}

\newcommand{\F}{\mathcal{F}}

\newcommand{\T}{{\cal T}}

\newcommand{\bra}{\langle}
\newcommand{\ket}{\rangle}

\newcommand{\tr}{{\rm tr}\,}

\title{}
\author{}

\begin{document}
\title{Entanglement  entropy of two disjoint intervals
from fusion algebra of twist fields}

\author{M. A. Rajabpour$^a$ and  F. Gliozzi$^b$  \\
~\\
$^a$ SISSA and INFN, Sezione di Trieste, via Bonomea 265,
34136 Trieste, Italy\\
$^b$ Dipartimento di Fisica Teorica, Universit\`a di Torino and\\ INFN,
Sezione di Torino, via P. Giuria, 1, 10125 Torino, Italy}

\maketitle

\begin{abstract}
We study the entanglement and R\'enyi entropies of two disjoint
intervals in minimal models of conformal field theory.
We use the conformal block expansion  and fusion rules of twist fields
to define a systematic expansion in the elliptic parameter of the trace
of the $n-$th power of the reduced density matrix. Keeping only the first few terms we obtain an  approximate expression that is easily analytically
continued to  $n\to1$, leading to an approximate formula for the entanglement entropy. These predictions  are checked against
some known exact results as well as against existing numerical data.
\end{abstract}
\section{Introduction}
 A useful quantity to study the phenomenon of quantum  entanglement
in extended  systems with many degrees of freedom is the von Neumann
or entanglement entropy. It is defined as follows. If a pure
quantum state $\vert\Psi\ket$ (typically the ground state)  can be
subdivided into two complementary subsystems $A$ and $B$ one can construct
the reduced density matrix
\eq
\rho_A=\tr_B\,\vert\Psi\ket\bra\Psi\vert
\en
by tracing over the degrees of freedom of $B$. The entanglement
entropy $S_A$ is simply the von Neumann entropy associated to $\rho_A$
\eq
S_A=-\tr\rho_A\ln \rho_A~.
\label{eentropy}
\en

The entanglement entropy has been extensively studied in low dimensional quantum systems as a new way to investigate the nature of quantum criticality
\cite{cw,hlw,vid,ch,jk,cc,cp}.
Several different calculations based on the conformal field theory (CFT)
describing the universal properties of the quantum phase transitions in 1+1
dimensional systems, like  spin chains, have shown that the entropy grows
logarithmically with the size $\ell$ of the subsystem $A$ as
\cite{hlw,vid,ch,jk,cc,RT}
\eq
S_A=\frac c3 \log\, \ell+k~,
\label{esc}
\en
where $c$ is the central charge of the CFT and $k$ is a non-universal constant
 related to the ultraviolet cutoff.

As an extension of the von Neumann entropy one considers the R\'enyi entropy, defined as
\eq
R_A^{(n)}=\frac{-1}{n-1}\log\tr \rho_A^n~.
\label{renyi}
\en
The von Neumann entropy can be reached in the limit $n\to1$.

The quantity $\tr \rho_A^n$
plays a major role in the replica approach to entanglement
entropy \cite{hlw,cc}. This method is based on the fact that, for integral $n$,
 $\tr \rho_A^n$  is the ratio $Z_n(A)/Z^n$, where $Z_n(A)$ is the partition function on a $n-$sheeted  Riemann surface, obtained by joining cyclically the $n$
sheets along region $A$, and $Z$ is the partition function of a single
sheet. In one-dimensional quantum systems the subsystem $A$ consists of one or
more disjoint intervals $A=A_1\cup A_2\cup\dots$ and
the trace $\tr\rho_A^n$ is proportional,
at criticality, to the $n-$th power of the correlation function of
local primary fields sitting on the end points $u_i,v_i$  of $A_i$
\eq
\tr \rho_A^n\propto \bra\T(u_1,\bar{u}_1)   \tilde{\T}(v_1,\bar{v}_1)\dots \ket^n~.
\label{rhobp}
\en
These primary fields have  conformal weight \cite{Knizhnik}
\eq
\Delta_n=\bar\Delta_n=\frac c{24}\left(1-\frac1{n^2}\right)
\label{delta}
\en
hence do not belong in general to the Kac table.
They can be considered as a special kind of twist fields, called
branch-point twist fields \cite{ccd}, because they are naturally related
to the branch points in the $n$-sheeted Riemann surface where
the system is defined.

When $A$ consists of a single interval of length $\ell$ the equation
(\ref{rhobp}) yields
\eq
\tr\rho_A^n=c_n\ell^{-\frac c6(n-1/n)},
\label{single}
\en
where $c_n$ is a non-universal constant. This expression can be easily
analytically continued to any real or complex $n$ and the limit
\eq
-\lim_{n\to1 }\partial_n\
tr\rho_A^n =S_A
\en
 gives at once Eq. (\ref{esc}).

When the subsystem $A$ consists of more than one interval, the analysis becomes more complicated and a complete description is still lacking. In the case of two disjoint intervals $A=A_1\cup A_2=[u_1,v_1]\cup[u_2,v_2]$ global conformal invariance \cite{BPZ} gives
\eq
\tr\rho_A^n=c_n^2\left(\frac{\vert u_1-u_2\vert\vert v_1-v_2\vert}
{\vert u_1-v_1\vert\vert u_2-v_2\vert\vert u_1-v_2\vert\vert u_2-v_1\vert}
\right)^{\frac c6(n-1/n)}\F_n(x),
\label{bpz}
\en
 where $x$ is the cross-ratio
\eq
x=\frac{(u_1-v_1)(u_2-v_2)}{(u_1-u_2)(v_1-v_2)}~.
\en
It was formerly assumed \cite{cc} $\F_n(x)\equiv1$ identically, but
it was subsequently realized \cite{CG,FPS} through analytical and numerical
means that this choice was incorrect. This observation generated in last
years an intense research work aimed at determining analytically and/or
numerically the function
$\F_n(x)$ \cite{CCT1,CH2,FFI,ACT,ATC2,IP,FC,Cal,JK2,MF,CCT}. In particular in
\cite{CCT} it has been shown that in any CFT  $\F_n(x)$  admits a small $x$
 expansion and the first few terms have been evaluated.

Our goal in this paper is to describe a general method to calculate the function $\F_n(x)$ in minimal models using conformal blocks and fusion rules of
twist fields. In section \ref{four point} we apply an idea of Zamolodchikov of expanding
the conformal blocks with respect to the elliptic variable $q$
(see next section for details). Since even very small $q$'s can be
related to large $x$'s, this expansion, even by just considering only
the first few terms, gives a good approximation for $\F_n(x)$ (see Eq.
(\ref{key formula})), that is easily analytically continued
to any $n$. This leads  to write an approximate expression for the
contribution of $\F_n$ to the von Neumann entropy
$\F_{VN}(x)=\lim_{n\to1}\partial_n\F_n(x)$, which fits nicely to the numerical
data of the critical Ising model. Our formula for $\F_n(x)$ also compares favorably with exact results at $n=2$ which are described in the last section.
An  appendix describes some properties of the two functions
$s_2(n,\alpha)$ and $s_4(n,\alpha)$ which give
the contribution  of the two spin and  four spin operators to
$\F_n(x)$.

\section{Four point function of twist fields and conformal block technique}
\label{four point}
In this section we show how one can calculate the four point
function of  twist operators on a sphere by using the structure
constants and conformal blocks. In the first subsection we introduce
the conformal block expansion of four point function and the
Zamolodchikov's elliptic recursion relation to calculate the
conformal blocks. In the second subsection we summarize the fusion
structure of the twist fields for different copies of conformal
field theories. In the last subsection, combining the results of the two
subsections, we propose some good approximate results for the R\'enyi
and Von Neumann entanglement entropy of minimal models.

\subsection{Zamolodchikov's recursion relation}

Using global conformal symmetry one can always write down any four
point function as
 \begin{eqnarray}\label{Four point function}
<\mathcal{O}_1(z_1,\bar{z}_1)\mathcal{O}_2(z_2,\bar{z}_2)\mathcal{O}_3(z_3,\bar{z}_3)\mathcal{O}_4(z_4,\bar{z}_4)>=
\hspace{4cm}\nonumber\\\Big{(}\frac{z_{13}z_{24}}{z_{14}z_{23}z_{12}z_{34}}\Big{)}^{2\bar{\Delta}_i}.\Big{(}\frac{\bar{z}_{13}\bar{z}_{24}}{\bar{z}_{14}
\bar{z}_{23}\bar{z}_{12}\bar{z}_{34}}\Big{)}^{2\Delta_i}\mathcal{F}_{34}^{12}(x,\bar{x}),\hspace{1cm}
\end{eqnarray}
where
\begin{eqnarray}\label{eta}
x=\frac{z_{12}z_{34}}{z_{13}z_{24}},\hspace{1cm}\bar{x}=\frac{\bar{z}_{12}\bar{z}_{34}}{\bar{z}_{13}\bar{z}_{24}},
\end{eqnarray}
is the cross ratio and $\Delta_i$ is the conformal weight of the
operator $\mathcal{O}_i(z_1,\bar{z}_1)$.

It is well-known that $\mathcal{F}_{34}^{12}(x,\bar{x})$ in
any conformal field theory can be expanded with respect to the
conformal blocks as \cite{BPZ}
\begin{eqnarray}\label{conformal block expansion}
\mathcal{F}_{34}^{12}(x,\bar{x})=\sum_{l}C_{12}^{l}C_{34}^{l}F(\tilde{c},\Delta_{l},\Delta_{i},x)F(\tilde{c},\bar{\Delta}_{l},\bar{\Delta}_{i},\tilde{x});
\end{eqnarray}
where $\tilde{c}$ is the central charge of the conformal field
theory\footnote{We reserve $c$ as the central charge of one copy of
minimal CFT in the next pages.},
$F(\tilde{c},\Delta_{l},\Delta_{i},x)$ is the conformal block,
$C_{12}^{l}$ is the structure constant and the indices $l$ indicate
the fusion channels. For those cases that at least one of the
$\mathcal{O}_i$'s has a null vector \footnote{In other words when at
least one of them is part of the Kac table.} one can usually write a
differential equation that $F(\tilde{c},\Delta_{l},\Delta_{i},x)$
satisfies. Although in limited cases one can find the conformal
blocks explicitly \cite{BPZ} it is usually very difficult to find a
compact formula for the conformal blocks. When none of the
$\mathcal{O}_i$'s has a null vector one can simply write an
expansion of the conformal block with respect to the cross ratio.
However, as was already noticed in \cite{BPZ}, this expansion
converges very slowly and so one needs to do numerical calculations
to get reasonable results. A recursion relation formula was found by
Al Zamolodchikov \cite{Zamol1} which is more suited for numerical
calculation, however still the convergence is slow and one can not
get interesting results by just taking the first few terms. Al
Zamolodchikov was able to find another recursive formula by
expanding the conformal blocks with respect to the elliptic variable
$q$; \cite{Zamol2,Zamol3}. Since even very small $q$'s can be
related to large $x$'s, this expansion, even by just taking few
  terms, gives very good approximation of the conformal blocks
 for large $x$'s. The Zamolodchikov's formula for the
conformal block has the following form

\begin{eqnarray}
F(\tilde{c},\Delta_{l},\Delta_{i},x)&=&(16q)^{\Delta_{l}-\frac{\tilde{c}-1}{24}}x^{\frac{\tilde{c}-1}{24}}(1-x)^{\frac{\tilde{c}-1}{24}
 }(\theta_{3}(\tau))^{\frac{\tilde{c}-1}{2}-4\delta}H(\tilde{c},\Delta_{l},\Delta_{i},q);\label{conformal block 2}\hspace{0.75cm}\\
H(\tilde{c},\Delta_{l},\Delta_{i},q)&=&1+\sum_{m,n}(16q)^{mn}\frac{R_{mn}(\tilde{c},\Delta_{i})
H(\tilde{c},\Delta_{m\,n}+mn,\Delta_{i},q)}{\Delta_{l}-\Delta_{m\,n}(\tilde{c})};\label{H}
\end{eqnarray}
where $\delta=\sum_{i=1}^4\Delta_{i}$  and
\begin{eqnarray}\label{kac table weights}
\Delta_{m\,n}(\tilde{c})&=&\frac{\tilde{c}-1}{24}+\frac{(\beta
m-\beta^{-1}n)^2}{4}.
\end{eqnarray}
The $q$ is the elliptic variable and has the following relation with
the cross ratio

\begin{eqnarray}\label{ q with respect to eta}
q=e^{i\pi \tau};\hspace{1cm} \tau=\frac{i K'(x)}{K(x)}=\frac{i
K(1-x)}{K(x)},\hspace{1cm}
x=\frac{\theta_{2}^{4}(\tau)}{\theta_{3}^{4}(\tau)},
\end{eqnarray}
where $K(x)$ is the complete elliptic integral of the first kind and
$\theta_{i}(\tau)$'s are the Jacobi elliptic functions defined as
\begin{eqnarray}\label{K}
K(x)=\frac{1}{2}\int_{0}^{1}\frac{dt}{\Big{(}t(1-t)(1-xt)\Big{)}^{1/2}},\\
\theta_{2}(\tau)=\sum_{n\in \mathbf{Z}}q^{(n+1/2)^2},\hspace{1cm}
\theta_{3}(\tau)=\sum_{n\in \mathbf{Z}}q^{n^2}.
\end{eqnarray}
$R_{m\,n}(c,\Delta_{i})$ has the following complicated form
\begin{eqnarray}\label{R}
R_{m\,n}(\tilde{c},\Delta_{i})=-\frac{1}{2}\prod'_{k,l}\frac{1}{\lambda_{kl}}\times\hspace{7cm}\nonumber\\\prod_{p,q}(\lambda_1+\lambda_2-\frac{\lambda_{pq}}{2})(-\lambda_1+\lambda_2-\frac{\lambda_{pq}}{2})(\lambda_3+\lambda_4-\frac{\lambda_{pq}}{2})(\lambda_3-\lambda_4-\frac{\lambda_{pq}}{2});
\end{eqnarray}
where
\begin{eqnarray}\label{notations}
\lambda_{pq}&=&p\beta-q\beta^{-1},\nonumber\\
\beta&=&\frac{1}{\sqrt{24}}\Big{(}(1-\tilde{c})^{1/2}+(25-\tilde{c})^{1/2}\Big{)},\\
\Delta_{i}&=&\frac{\tilde{c}-1}{24}+\lambda_{i}^2,\nonumber
\end{eqnarray}
and the products are taken over the following sets:
\begin{eqnarray}\label{range of nmkl}
p&=&-m+1,-m+3,...,m-3,m-1,\nonumber\\
q&=&-n+1,-n+3,...,n-3,n-1,\nonumber\\
k&=&-m+1,-m+2,...,m-1,m,\nonumber\\
l&=&-n+1,-n+2,...,n-1,n.\nonumber
\end{eqnarray}
 The prime on the symbol of the first product in (\ref{R}) means that the factors with $(k,l)=(0,0)$ and $(m,n)$ must be omitted.

 To have an idea about the expansion we write the first few terms of $H(c,\Delta_{l},\Delta_{i},q)$
 as
\begin{eqnarray}\label{H expansion}
H(\tilde{c},\Delta_{l},\Delta_{i},q)&=&1+\sum_{k=1}h_k(\tilde{c},\Delta_{l},\Delta_{i}) (16q)^k,\\
h_1(\tilde{c},\Delta_{l},\Delta_{i})&=&\frac{R_{11}(\tilde{c},\Delta_{i})}{\Delta_l},\\
h_2(\tilde{c},\Delta_{l},\Delta_{i})&=&\Big{(}\frac{R_{11}^2(\tilde{c},\Delta_{i})}{\Delta_l-\Delta_{11}}+
\frac{R_{12}(\tilde{c},\Delta_{i})}{\Delta_l-\Delta_{12}}+\frac{R_{21}(\tilde{c},\Delta_{i})}{\Delta_l-\Delta_{21}}\Big{)}.
\end{eqnarray}

There are some comments in order:
\begin{enumerate}
\item Looking to the equation (\ref{ q with respect to eta}) one can easily
see that the expansion with respect to $q$ is converging quite
faster than the expansion with respect to $x$.
\item The expansion (\ref{H}) has a singularity
whenever $\Delta_{l}$ is  part of the Kac table. This means that the
expansion is not useful in calculating the four point function of
the operators in the Kac table.
\item The expansion (\ref{conformal block 2}) was derived in
\cite{Zamol3} for $\tilde{c}\leq1$ and $\tilde{c}\geq25$. The
technique is based on using  the expansion of classical block for
large central charges and large conformal weights. In calculating
the classical block one needs to study the behavior of the five
point function ( the original four primary operators and an operator
with the second level null vector). The second level null vector is
just present in conformal field theories with $\tilde{c}\leq1$ and
$\tilde{c}\geq25$ . Although in deriving the classical behavior of
conformal block one needs to use second level null vector, it is
quite likely that the final result be correct for generic central
charges by analytical continuation. Another important thing to
notice in this direction is related to  possible complex numbers (
with non-zero imaginary part) that can appear in (\ref{conformal
block 2}) for $1<\tilde{c}<25$. It is easy to see that
$\Delta_{m\,n}(\tilde{c})$ is not real in the region
$1<\tilde{c}<25$. Using (\ref{R}) it is not difficult to show that
in this region $R_{mn}(\tilde{c},\Delta_{i})$ has also imaginary
part. However, one can show that in the first few terms of the
expansion  the imaginary parts all disappear and we have a real
number. We assume this is true for all the terms. Since in this work
we deal with $n$ copies of CFT's we will usually have $1\leq
\tilde{c}<25$, so we will assume that Zamolodchikov's formula is
correct also in this region.
\item Based on the associativity of the conformal algebra one
expects
crossing symmetry\footnote{This is also related to the modular
invariance of the theory.} for the four point functions of primary
operators in any CFT. In other words
\begin{eqnarray}\label{crossing symmetry}
\mathcal{F}_{34}^{12}(1-x,1-\bar{x})=\mathcal{F}_{34}^{12}(x,\bar{x}).
\end{eqnarray}
It seems that it is a very difficult task to show analytically that
the  Zamolodchikov's formula satisfies this relation. However, it is
not difficult to show numerically that by keeping the first few terms of
the formula (\ref{conformal block 2}) the equality (\ref{crossing
symmetry}) is approximately valid.

\end{enumerate}

\subsection{Fusion rules and structure constants of twist operators}

To calculate the four point function of the twist fields by using
conformal block technique one needs to know the fusion structure of
two twist fields. The OPE of two twist fields can produce many
different operators, in case of minimal models for generic $n$ we do
not know how to write all the primary operators that can be
produced, however, we know surely which operators will be there. For
simplicity consider unitary minimal models $\mathcal{M}(p,p+1)$ with
the central charge $\tilde{c}=c$ and primary operators $\phi_{rt}$
with conformal weights
\begin{eqnarray}\label{minimal models}
\Delta_{rt}&=&\frac{(r(p+1)-tp)^2-1}{4p(p+1)};\hspace{1cm}1\leq r \leq p;\hspace{0.5cm} 1\leq t \leq p-1\\
c&=&1-\frac{6}{p(p+1)}.
\end{eqnarray}
When we have $n$ copies of a CFT we can label the primary operators
with the upper indices $i$ like $\phi_{rt}^i$; where $i=1,2,...,n$.
One can easily show that the OPE of two twist fields will not
produce one copy of $\phi_{rt}^i$. However, $\phi_{rt}^i\phi_{rt}^j$
which is made of two copies of primary operators, with the conformal
weight $2\Delta_{rt}$, always appears after the fusion of two twist
operators. In general when we have $n\geq2$ the combination of
$m(1<m\leq n)$ copies of the primary operators, with the conformal
weight $m\Delta_{rt}$, will always appear after the fusion. For
$n=2$ it was shown in \cite{CSS} that these operators exhaust the
fusion structure but for $n>2$ there is no conclusive classification
for the fusion structure. In our approximate method this will not be
a big problem as far as we consider not very big $n$'s as we will
see soon. The next step is calculating the structure constants. The
structure constants for small $m$'s were already calculated in
\cite{Headrick} and \cite{CCT}. For example for $m=2$ one can show
\begin{eqnarray}\label{structure constants generic case}
C^2_{l=rt}&=&\left(\frac{1}{4n^2}\right)^{4\Delta_{rt}}s_2(n,4\Delta_{rs}),\\
s_2(n,\alpha)&=&\frac{n}{2}\sum_{j=1}^{n-1}\frac{1}{\Big{(}\sin(\frac{\pi
j}{n})\Big{)}^{2\alpha}}.
\end{eqnarray}
In principle calculating the structure constants for the operators
with conformal weight $m\Delta_{l=rt}$ is related to the calculation
of the $m$-point function of the operator $\phi_{rt}$. The
calculation is practically doable just for  $m\leq4$ and small $r$
and $t$ ( to be precise even for $m=4$ with $r>2$ and $t=1$ the
calculation is very cumbersome). Fortunately, as we will show, most of
the terms with $m>2$ give a very small contribution and so one can
ignore them in the first approximation. In addition there are also
some cases where the structure constants for some fields with $m>2$
is zero, this is the consequence of the zero $m$-point function of
the corresponding primary operators. For example, since the three
point function of the spin operators and the energy operators in the
Ising model is zero, the structure constants for three copies of the
spin operator and the energy operator are always zero, this is the
case also for all odd $m$'s in the Ising model.

\begin{figure} [htb]
\centering
\includegraphics[width=1\textwidth]{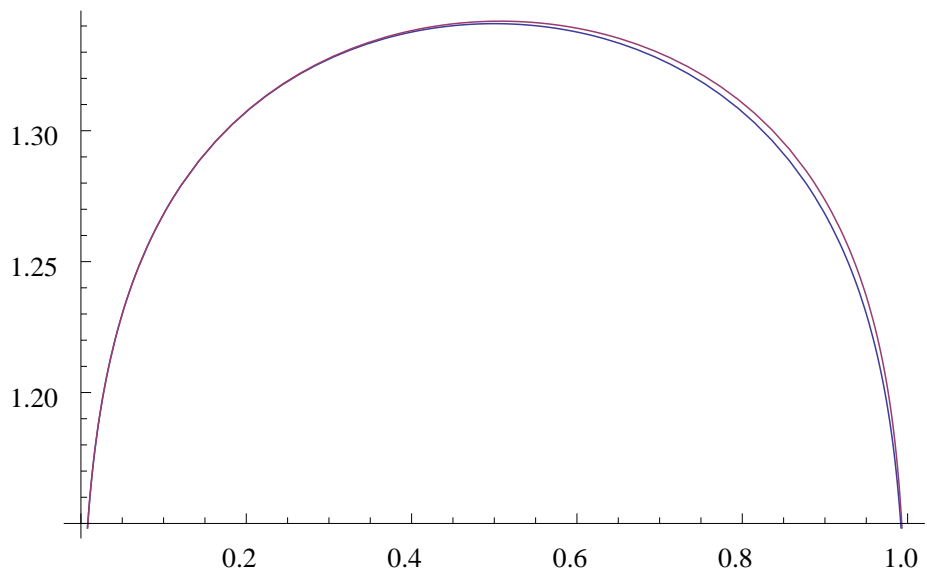}
\caption{$\mathcal{F}_{2}(x)$ for the Ising model. The blue line
is the exact result obtained in \cite{ACT} and the red one comes
from the equation (\ref{key formula}).} \label{Figure:1}
\end{figure}

\begin{figure} [htb]
\centering
\includegraphics[width=1\textwidth]{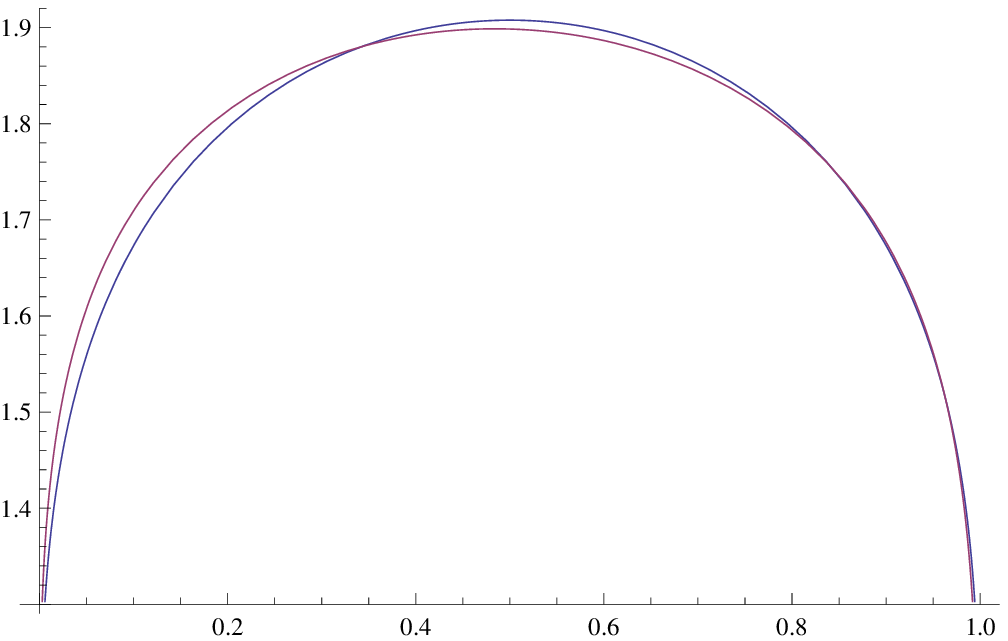}
\caption{$\mathcal{F}_{3}(x)$ for the Ising model. The blue line
is the exact result obtained in \cite{CCT} and the red one comes
from the equation (\ref{key formula}).} \label{Figure:2}
\end{figure}

\begin{figure} [htb]
\centering
\includegraphics[width=.8\textwidth,angle=-90]{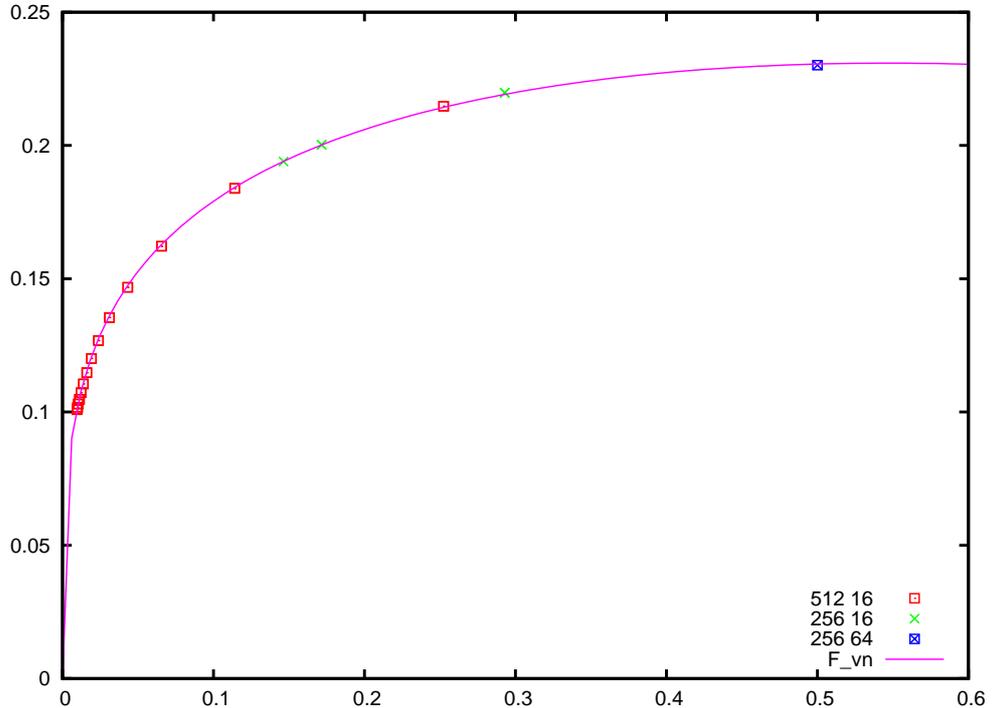}
\caption{The function $\mathcal{F}_{VN}(x)$ for the Ising model as a function of the cross-ratio $x$. The solid line is the plot
of Eq. (\ref{Von Neumann Ising}), while the data are taken form Ref. \cite{ACT}. The two integers in
the legend are respectively the size $L$ of the system and the size of the interval $\ell$ in units of lattice spacing.} \label{Figure:3}
\end{figure}

\subsection{R\'enyi entropy}

In this subsection by using the results of the last two subsections
we give an approximate formula for the R\'enyi entropy. Using the
formula (\ref{conformal block 2}) with $\tilde{c}=nc$ and
$\delta=\frac{nc}{6}(1-\frac{1}{n^2})$ one can write the following
formula

\begin{eqnarray}\label{four point function of twist field}
\mathcal{F}_{n}(x)\sim\frac{1}{(\theta_{3}(\tau))^{1+\frac{nc}{3}-\frac{4c}{3n}}}
\left(\frac{x(1-x)}{16q}\right)^{\frac{nc-1}{12}}
\sum_{l}C^2_l(16q)^{2\Delta_{l}}+...,
\end{eqnarray}
where we take $\mathcal{F}_{34}^{12}=\mathcal{F}_{n}$. Since the
above series is converging very fast, one can just take
$H(\tilde{c},\Delta_{l},\Delta_{i},q)\approx 1$ and also consider
just the contribution of the field with the minimal conformal weight
$\Delta_{min}$, then
\begin{eqnarray}\label{key formula}
\mathcal{F}_{n}(x)\sim\frac{1}{(\theta_{3}(\tau))^{1+\frac{nc}{3}-\frac{4c}{3n}}}
\left(\frac{x(1-x)}{16q}\right)^{\frac{nc-1}{12}}
\Big{(}1+s_2(n)\left(\frac{4q}{n^2}\right)^{2\Delta_{min}}+...\Big{)},
\end{eqnarray}
where we put $s_2(n)\equiv s_2(n,2\Delta_{min})$.
Notice that $\Delta_{min}$ in the above formula is  two times bigger
than the minimum conformal weight in the minimal models in
(\ref{minimal models}). To see how good is our approximation we 
discuss Ising model as the simplest unitary minimal model.

For Ising model and  $n=2$ the fusion rule is

\begin{eqnarray}\label{Ising fusion}
\mathcal{T}_2(z_1,\bar{z}_1)\bar{\mathcal{T}}_2(z_2,\bar{z}_2)=1+\sigma^1\sigma^2+\epsilon^1\epsilon^2;
\end{eqnarray}
where $\epsilon^i$  and $\sigma^i$ are the energy  and spin
operators of the $i$th copy respectively. In this case
$\Delta_{min}=\frac{1}{8}$ is the conformal weight of the two copies
of the spin operator. In Fig. \ref{Figure:1} we compare the equation (\ref{key
formula}) with the exact result derived in \cite{ACT}. The
excellent agreement is not a special feature of the critical Ising model.
Actually we will verify in section \ref{tworeplicas}, where we describe the
exact form of $\F_2(x)$, that   Eq.(\ref{four point function of twist field})
for $n=2$ is an excellent approximation for any conformal  model.

For  $n=3$ part of the fusion rule is
\begin{eqnarray}\label{Ising fusion 3}
\mathcal{T}_3(z_1,\bar{z}_1)\bar{\mathcal{T}}_3(z_2,\bar{z}_2)=1+(\sigma^1\sigma^2+perm)+(\epsilon^1\epsilon^2+perm)+(\sigma^1\sigma^2\epsilon^3+perm)+...,
\end{eqnarray}
where "$perm$" means that we need to consider all the permutations of the three indices $i=1,2,3$.
The dots consider the primary operators that can appear by combining
the descendants of different copies of the primary operators
\cite{Cardy}. Since the conformal weight of the operator
$\sigma^i\sigma^j\epsilon^k$ is quite bigger than the conformal
weight of the operator $\sigma^i\sigma^j$ we do expect that the
dominant term be still the contribution of $\sigma^i\sigma^j$. In
Fig \ref{Figure:2} we compare the result coming from the equation (\ref{key
formula}) with the exact result obtained in \cite{CCT}. Although we
ignored many contributions, the result is still satisfactory. In
principle the result should be better for small $x$'s than the
large ones, however, for unknown reason our formula works better for
large $x$'s. It is worth mentioning that if one  wants to calculate
the next terms just considering the terms like
$\sigma^1\sigma^2\epsilon^3$ is not enough. The reason is that
the higher order terms in the expansion of
$H(\tilde{c},\Delta_{l},\Delta_{i},q)$ for two copies of a primary
field could have equal or bigger contribution. Although this is not
the case for the Ising model, it simply shows how much it could be
complicated to go to the higher levels and keep the calculations
consistent. For future use we also give the first few terms of the
result for $n>3$. In this case the most dominant term after the
contribution of the two spin operators is the contribution of four
spin operators with the conformal weight $\frac{1}{4}$. Since the
power of $q$ in this case in equation (\ref{four point function of
twist field}) is one half, one expects an important contribution from
this term. One can write
\begin{eqnarray}\label{key formula Ising}
\mathcal{F}_{n}(x)\sim\frac{1}
{(\theta_{3}(\tau))^{1+\frac{n}{6}-\frac{2}{3n}}}\left(\frac{x(1-x)}{16q}\right)^{\frac{n-2}{24}}
\Big{(}1+s_2(n)\left(\frac{4q}{n^2}\right)^{\frac{1}{4}}+s_4(n)
\left(\frac{4q}{n^2}\right)^{\frac{1}{2}}+...\Big{)},
\end{eqnarray}
where $s_4(n)\equiv s_4(n,2\Delta_{min})$ is calculated in \cite{CCT} and has the following form
\begin{eqnarray}\label{s_4n}
s_4(n,\alpha)=\frac{n}{4}\sum_{_{1\leq j_2<j_3<j_4\leq
n-1}}\left(\frac{\sin(\pi j_{42}/n)\sin(\pi j_{31}/n)}{\sin(\pi
j_{21}/n)\sin(\pi j_{43}/n)\sin(\pi j_{41}/n)\sin(\pi j_{32}/n)}\right)^{2\alpha},
\end{eqnarray}
where $j_{kl}=j_k-j_l$ and $j_1=1$.  Some properties of $s_2(n,\alpha)$ and $s_4(n,\alpha)$
are discussed in the Appendix.

\subsection{Von Neumann  entropy}

Since the equations (\ref{key formula}) and (\ref{key formula
Ising}) are  simple functions of the number of replicas $n$ one can
simply calculate the von Neumann entanglement entropy by just
differentiating them with respect to $n$ and evaluating them at $n=1$.
Then for Ising model we have
\begin{eqnarray}\label{Von Neumann Ising}
\frac{d\mathcal{F}_{n}(x)}{d n}|_{n=1}\sim
\theta_3^{\frac{-1}{2}}(q)\left(\frac{x(1-x)}{16q}\right)^{\frac{-1}{24}}\Big{(}-\frac{5}{6}\log\theta_3(q) + \frac{1}{48}\log\left(\frac{x(1-x)}{16q}\right)+\nonumber\\
s'_2(1)\left(4q\right)^{\frac{1}{4}}+s'_4(1)\left(4q\right)^{\frac{1}{2}}\Big{)}+...,\hspace{3cm},
\end{eqnarray}
where $s'_2(1)$ is known from the work in \cite{CCT} and for generic
conformal weight has the following form
\begin{eqnarray}\label{s2n n=1}
s'_2[1,2\Delta_{min}]=+\frac{\sqrt{\pi}
\Gamma(1+2\Delta_{min})}{4\Gamma(2\Delta_{min}+\frac{3}{2})}.
\end{eqnarray}
Calculation of $s'_4(1,\frac14)$ is a very difficult task, we find an
approximate value for this quantity by using a numerical method
described in the Appendix which yields $s'_4(1)\simeq0.126(3)$.

We compare Eq.(\ref{Von Neumann Ising}) with the numerical data
\footnote{We thank the authors of Ref. \cite{ACT} for providing us
with their numerical data.}  of Ref. \cite{ACT}, based on a tree
tensor network algorithm (TTN) \cite{TTN} applied to a critical one
dimensional quantum spin chain in transverse field, corresponding to
the CFT minimal model with $c=\frac12$. This algorithm gives the
full spectrum of the reduced density matrix $\rho_A$. From this  the
moments $\tr \rho_A^n$ and the entanglement entropy $S_A$  are
easily evaluated. The data of \cite{ACT} are taken for a periodic
system of length $L$ with a subsystem  $A=[v_1-u_1]\cup [v_2-u_2]$
composed of two disjoint intervals of identical length
$\ell=v_1-u_2=v_2-u_2$ at distance $r$. The cross-ratio $x$ is given
by \eq x=\left(\frac{\sin\pi \ell/L}{\sin\pi(\ell+r)/L}\right)^2.
\en The ratio $\rho_{A}^n/\rho_\ell^{2n}$ eliminates the non
universal constant $c_n$ of Eq. (\ref{bpz}) and allows to evaluate,
up to obvious factors, the universal function $\F_n(x)$  and its
contribution $F_{VN}=\partial_n\F_{n}(x)\vert_{n=1}$ to the von
Neumann entropy. In Fig. \ref{Figure:3} we plot the function
$\F_{VN}(x)$ as well as the TTN numerical data, finding a perfect
agreement. It is interesting to note that if we use the quantity
$s'_4[1,\frac14]$ as a free parameter to fit these data we find a
value which is consistent, within the errors, with the value
estimated in the Appendix in a completely different context.

\section{The special case of two replicas}
\label{tworeplicas}
The case $n=2$ is very interesting and instructive.
First, note that
Eq. (\ref{four point function of twist field})  simplifies dramatically
because the  structure constants become simply
\eq
C^2_{l=rt}=\left(\frac{1}{16}\right)^{4\Delta_{rt}}~,
\en
thus
\eq
\mathcal{F}_{2}(x)\sim\frac{1}{\theta_{3}(\tau)}
\left(\frac{x(1-x)}{16q}\right)^{\frac{c}6-\frac1{12}}
\sum_{l}q^{2\Delta_{l}}+....
\label{F2}
\en
Moreover
in this case the four-point function of the twist fields is directly related
to the formulation of the CFT on a two-sheeted Riemann surface with two cuts.
This  is conformally equivalent to a torus whose modular parameter $\tau$ is
related to the position of the branch points
by Eq.(\ref{ q with respect to eta}). As a consequence it is expected that the torus partition function $Z(\tau)$ of the CFT  should show up as a
factor of $\F_2$  \cite{Lunin,Headrick,ACT}.

In this section we would like to reconstruct this relationship within
our approach by presenting evidence that a factor of
Eq. (\ref{F2}) is just a truncated expansion
 of the partition function $Z(\tau)$.
To this aim, we resort to the following useful formulas
\eq
\left[x(1-x)\right]^{\frac1{12}}=\frac{\left(\theta_2(\tau)\theta_3(\tau)
\theta_4(\tau)\right)^{\frac13}}{\theta_3(\tau)}~,~~
\theta_2(\tau)\theta_3(\tau)\theta_4(\tau)=2\eta(\tau)^3.
\en
The Jacobi theta functions
$\theta_2(\tau)$ and $\theta_3(\tau)$ have been already defined in
(\ref{ q with respect to eta})
and similarly we have $\theta_4(\tau)=\sum_{n\in \mathbf{Z}}(-1)^nq^{n^2}$;
$\eta(\tau)$ is the Dedekind eta function, defined as
\eq
\eta(\tau)=q^{\frac1{12}}\prod_{n=1}^\infty(1-q^{2n})~;~~q=e^{i\pi\tau}\,.
\en

These relations, once inserted in Eq. (\ref{F2}), yield
\eq
\F_2(x)\sim \left[\frac{x(1-x)}{16}\right]^\frac c6\sum_{r\,t}
\vert e^{i2\pi\tau(\Delta_{r\,t}-c/24)}\vert^2
\en
where we restored the notation of Eq. (\ref{minimal models}) and the sum is
made over the entries of the Kac table of a given minimal model.
One recognizes at once that such a sum is just the truncation of the first few terms of the partition function
\eq
Z(\tau)=\sum_{r\,t}
\vert e^{2i\pi\tau(\Delta_{r\,t}-c/24)}\vert^2+{\rm higher~order~ terms},
\en
thus one is led to conjecture that the exact form of $\F_2(x)$ is simply
\eq
\F_2(x)= \left[\frac{x(1-x)}{16}\right]^\frac c6 Z(\tau)
\equiv \left\vert\frac{\eta(\tau)}{\theta_3(\tau)}\right\vert^{2c}Z(\tau)
\label{Ftwo}
\en
Actually  this formula coincides for $c=1$ with
the exact result for compactified boson found long time ago
\cite{Zamol4,DFMS,Lunin} and for $c=\frac12$ with the exact result
of the  critical Ising model obtained in \cite{ACT}.

\section{Conclusion}

In this paper we found  approximate formula (\ref{key formula}) for
the R\'enyi entropy ( with arbitrary $n$) of two disjoint intervals
of generic minimal conformal field theory. The idea was based on
using the elliptic expansion of the conformal blocks appearing in
the four point function of the twist fields. Since our formula had a
simple relation with  $n$, the number of replicas, we were able to
do the analytical continuation $n\rightarrow1$ easily and found the
von Neumann entanglement entropy. The von Neumann entropy of two
disjoint intervals in the Ising model was investigated as a
benchmark and we found excellent agreement between our formula
(\ref{Von Neumann Ising}) and the available numerical results.

Using the perturbative  expansion of the four point function of the
twist fields in the $n=2$ case  we were also able to find a new way to
connect explicitly the four point function of the twist fields to
the partition function of the conformal field theory on the torus.

Although our perturbative results give rather accurate formulas, even
by taking just the first term of the expansion, for the R\'enyi
entropy and  von Neumann entropy, it is still tempting to
understand how the results can be improved by going to the higher
levels. In particular, calculating the von Neumann entropy of more
complicated minimal models and checking the results with numerical
calculations can shed some light on the possible generalization of
our results. Since going to the higher levels of the perturbation is
possible just by calculating the structure constants appearing in
the OPE of two twist fields - which is itself related to the $k$ point
functions of the primary operators - we
are faced with the old, unsolved problem of conformal field
theory of calculating  correlation functions for arbitrary number 
of fields.

\vskip  .5 cm
\begin{flushleft}
{\bf Acknowledgments}
\end{flushleft}
FG would like to thank Pasquale Calabrese and Luca Tagliacozzo for fruitful
correspondence. MAR thanks John Cardy for many fruitful discussions
and comments.

\appendix
\section{ The functions $s_2(n,\alpha)$ and $s_4(n,\alpha)$}
\renewcommand{\theequation}{A.\arabic{equation}}
\setcounter{equation}{0}
In this appendix we deal with the functions
 $s_2(n,\alpha)$ and $s_4(n,\alpha)$, where $\alpha$ is related with the 
conformal weight by $\Delta=\alpha/4$.

The function $s_2(n,\alpha)$ has a
simple zero at $n=1$ while $s_4(n,\alpha)$ has
simple zeros at $n=1,2,3$. For integer $n$ both functions have moreover a factor
$n$ produced by the translation invariance on the sum of the indices $j_i$
so we may assume that both functions have also a zero  at $n=0$.
This is in particular true for  $s_2[n,k]$  for integral $k$,
where it has been shown in two different contexts and two different ways
\cite{Billo:1996gx,Headrick}  that $s_2[n,k]$
is a polynomial of degree $2k$.
The first few polynomials of this set are
\eq
s_2(n,1)=\frac n2\sum_{j=1}^{n-1}\frac1{\sin^2\frac{\pi j}n}=n\left(
\frac{n^2}6-\frac16\right),
\en
(this expression has also been found in \cite{CCT1})
\eq
s_2(n,2)=\frac n2\sum_{j=1}^{n-1}\frac1{\sin^4\frac{\pi j}n}= n
\left(\frac{n^4}{90}+
\frac{n^2}{9}-\frac{11}{90}\right),
\en
\eq
s_2(n,3)=\frac n2\sum_{j=1}^{n-1}\frac1{\sin^6\frac{\pi j}n}=n\left(\frac{n^6}{945}
+\frac{n^4}{90}+
\frac{4\,n^2}{45}-\frac{191}{1890}\right),
\en
\eq
s_2(n,4)=\frac n2\sum_{j=1}^{n-1}
\frac1{\sin^8\frac{\pi j}n}=n\left(\frac{n^8}{9450}
+\frac{4\, n^6}{2835}+\frac{7\,n^4}{675}+
\frac{8\,n^2}{105}-\frac{2497}{28350}\right)~.
\en
It is easy to verify in these cases the general formula of Ref. \cite{CCT}
\eq
s_2'(n,\alpha)_{n=1}=\frac{\sqrt{\pi}\,\Gamma(1+\alpha)}{4\,
\Gamma(\frac32+\alpha)}
=\frac14\,B\left(\frac12,1+\alpha\right),
\en
where $B(x,y)$ is the Euler Beta function. In particular, the contribution of the operator energy in critical Ising model is $s_2'(n,2)_{n=1}=\frac4{15}$
and one could add this term to Eq.(\ref{Von Neumann Ising}) to further improve
the formula for the von Neumann entropy.

Assuming that $s_2(z,\alpha)$ and $s_4(z,\alpha)$ are analytic functions on the complex plane $z$, they admit the following power expansions
\eq
s_2(z,\alpha)=z(z-1)\left(c_0+c_1z+c_2z^2+c_3 z^3+\dots\right),
\label{stwo}
\en
\eq
s_4(z,\alpha)=z(z-1)(z-2)(z-3)\left(d_0+d_1z+d_2z^2+d_3 z^3+\dots\right).
\label{sfour}
\en
If the two series expansions in the last parenthesis are rapidly convergent
series we have, approximately,
\eq
s_2'(1,\alpha)\simeq c_0+c_1+c_2+c_3,
\en
and similarly
\eq
s_4'(1,\alpha)\simeq 2\left(d_0+d_1+d_2+d_3\right)~.
\en
We fitted $c_0,c_1,c_2,c_3$ using the exactly calculable data
 for $s_2(n,\frac14)$ for integer $n=2,3,\dots 20$
and similarly $d_0,d_1,d_2,d_3$;
 $s_4(n,\frac14)$ for integer $n=4,5,\dots 20$
and verified in both cases that the coefficients $c_i$ and $d_i$ are rapidly decreasing. In this way $s_2'(1,\frac14)$ turns out to be compatible with the exact value. In the same way we estimated
\eq
s_4'(1,\frac14)\simeq 0.126(3)~,
\en
which is the value we put in Eq. (\ref{Von Neumann Ising}); it turns out to
fit nicely the numerical data as shown in Fig. \ref{Figure:3}. Using the fitted parameters $c_i$ and $d_i$ one could also study the analytic continuation of
$\F_n(x)$ to $\F_z(x)$, where $z$ is the complex plane of replicas
\cite{arXiv:0910.3003} simply by inserting the two functions (\ref{stwo})
and (\ref{sfour}) in Eq. (\ref{key formula}).

\end{document}